\documentclass[aps,prl,twocolumn,showpacs,superscriptaddress,footinbib]{revtex4-1} 

\usepackage[english]{babel}
\usepackage{graphicx}
\usepackage{dcolumn}
\usepackage{bm}
\usepackage{bbm}
\usepackage[utf8]{inputenc}
\usepackage{amssymb}
\usepackage{amsmath}
\usepackage{bbold}
\usepackage{pstricks}
\usepackage{slashed}
\usepackage{braket}
\usepackage{hyperref}
\usepackage{natbib}
\usepackage{float}
\usepackage{verbatim}
\usepackage{siunitx}
\usepackage{lipsum}
\usepackage{slashed}
\usepackage{ulem}

\usepackage[left=1.85cm,right=1.85cm,top=1.85cm,bottom=1.85cm]{geometry}

\definecolor{mygreen}{rgb}{0,0.5,0}
\definecolor{myblue}{rgb}{0,0,0.75}
\definecolor{mymagenta}{cmyk}{0,1,0,0.12}
\definecolor{mygray}{rgb}{0.5,0.5,0.5}

\newcommand{\Fig}[1]{Fig.~\ref{#1}}

\usepackage{umoline}

\begin{document}

\title{Scalable cold-atom quantum simulator for two-dimensional QED}

\author{R. Ott}
\email[]{ott@thphys.uni-heidelberg.de}
\affiliation{Heidelberg University, Institut f\"{u}r Theoretische Physik, Philosophenweg 16, 69120 Heidelberg, Germany}

\author{T. V. Zache}
\affiliation{Heidelberg University, Institut f\"{u}r Theoretische Physik, Philosophenweg 16, 69120 Heidelberg, Germany}
\affiliation{Center for Quantum Physics, University of Innsbruck, 6020 Innsbruck, Austria}
\affiliation{Institute for Quantum Optics and Quantum Information of the Austrian Academy of Sciences, 6020 Innsbruck, Austria}

\author{F. Jendrzejewski}
\affiliation{Heidelberg University, Kirchhoff Institute for Physics, Im Neuenheimer Feld 226, 69120 Heidelberg, Germany}

\author{J. Berges}
\affiliation{Heidelberg University, Institut f\"{u}r Theoretische Physik, Philosophenweg 16, 69120 Heidelberg, Germany}

%\date{\today}
    
\begin{abstract}
\vskip 0.4cm
We propose a scalable analog quantum simulator for quantum electrodynamics (QED) in two spatial dimensions. The setup for the U$(1)$ lattice gauge field theory employs inter-species spin-changing collisions in an ultra-cold atomic mixture trapped in an optical lattice. Building on the previous one-dimensional implementation scheme of Ref.~\cite{mil2020scalable} we engineer spatial plaquette terms for magnetic fields, thus solving a major obstacle towards experimental realizations of realistic gauge theories in higher dimensions. We apply our approach to pure gauge theory and discuss how the phenomenon of confinement of electric charges can be described by the quantum simulator.  	
\end{abstract}
\maketitle

 \textit{Introduction.} There are strong efforts to quantum simulate gauge theories such as quantum electrodynamics (QED) using various platforms including cold atomic gases \cite{yang2020observation,mil2020scalable,schweizer2019floquet}, Rydberg atoms \cite{bernien2017probing}, trapped ions \cite{martinez2016real,kokail2019self} or superconducting qubits \cite{klco2018quantum}. Much progress has been achieved in one spatial dimension, and first experimental realizations of small systems in two dimensions have been shown \cite{yang2017,klco20202,yamamoto2020real}. However, scalable experimental implementations of two- and higher-dimensional setups are still elusive. Here the main challenge is the efficient construction of gauge invariant plaquette terms of the lattice field theory, corresponding to magnetic field interactions. Their interplay with electric fields is crucial even for basic phenomena, like the propagation of transversely polarized photons, not present in one dimension. 

Gauge field dynamics in two or more spatial dimensions can give also access to one of the most intriguing phenomena: confinement. The most prominent example is the confinement of quarks into colorless hadrons, forming the basis for nuclear matter in quantum chromodynamics \cite{montvay1997quantum}. Another example is confinement of electric charges in compact quantum electrodynamics~\cite{polyakov}.
%In compact quantum electrodynamics one finds confinement of \emph{electric} charges. 
Its description can be a difficult problem for classical computational techniques and quantum simulators promise important progress in our abilities to address equilibrium as well as dynamical properties out of equilibrium from first principles.

So far, proposals to engineer the magnetic field interactions using atoms in optical lattices rely mainly on perturbative constructions \cite{zohar2011confinement,zohar2012simulating,gonzalez2017quantum,kasamatsu2013atomic,buchler2005atomic,paredes2008minimum}. Other approaches, using Rydberg atoms in optical tweezers \cite{CeliPRX}, two-dimensional arrays of superconducting qubits \cite{marcos2014two,brennen2016loops}, Floquet engineering in optical lattices \cite{barbiero2019coupling}, or universal quantum computers \cite{haase2020resource,paulson2020towards}, currently focus on gauge theories with small local Hilbert spaces, which are typically unable to describe important phenomena with large fluctuations of electromagnetic fields.

 %effective spin systems with small spin representations rather than QED. 

Here, we propose a new scheme to quantum simulate the gauge fields of compact QED in two spatial dimensions using a two-species mixture of ultra-cold spinor Bose gases trapped in an optical lattice. We represent electric fields by site occupations on a lattice \cite{yang2016analog,zohar2011confinement,kasamatsu2013atomic} and the electric field energy results from on-site interactions.  The engineering of gauge invariant plaquette terms is based on hetero-nuclear atomic collisions similar to those successfully demonstrated in the recent experiment of Ref.~\cite{mil2020scalable}. Neighboring unit plaquettes are connected by forming superpositions of the atomic clouds. This can be realized via resonant microwave dressing or laser-assisted tunneling. We exploit that our cold-atom system approaches QED by putting Bose-Einstein condensates (BECs) with large numbers of atoms on the lattice \cite{kasper2017implementing, zache2018quantum, zohar2013quantum}, rather than single atoms \cite{zohar2011confinement}.  The high tunability of the proposed approach gives access to a wide range of coupling strengths, including regimes of QED in which perturbative treatments are poorly~controlled. 

In our setup, confinement will be directly detectable as a finite change in occupation number on the links that connect two static charges, even for increasing distance $D\rightarrow \infty$ between the charges. 
As an illustration, we discuss the emergence of confinement in compact QED in the weak coupling regime, using a variational ansatz for the ground state wave function \cite{drell1979quantum}. We explain that the phenomenon of confinement will be robustly observable for the proposed cold-atom implementation against experimental imperfections due to finite boson numbers and unwanted inter-species density interactions of the~atoms.

\begin{figure}
	\flushleft
	\includegraphics[width=0.48\textwidth]{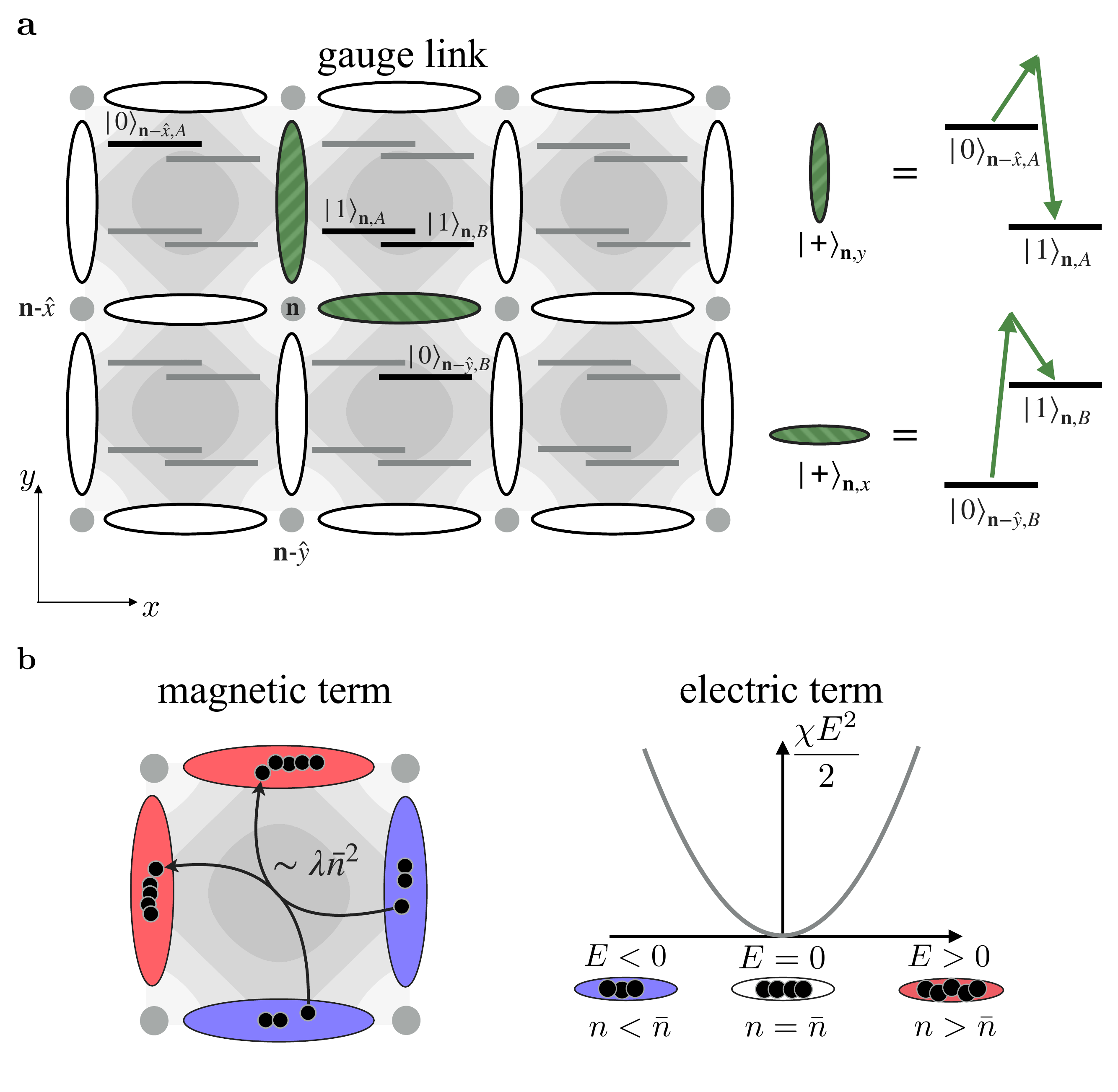}
	\caption{{\bf (a)} Sketch of the implementation scheme. Atomic clouds are trapped in the minima of the optical lattice at the center of the plaquettes (dark gray shaded regions). The gauge field degrees of freedom (ellipses) are formed by creating superpositions of the four hyperfine levels in each well with neighboring clouds along the $x$- or $y$-axis, respectively.
	{\bf (b)}~Magnetic plaquette interactions of four neighboring links are formed by atomic scatterings within the minima of the optical lattice (left). The electric field energy is given by homo-nuclear density interactions forming an on-site potential (right). The electric field strength is represented by the atom number on each link relative to a background number~$\bar{n}$.}
	\label{fig:1}
\end{figure}

\textit{Lattice gauge field theory.} In the Hamiltonian formulation of compact QED we introduce electric fields and gauge links on a two-dimensional spatial lattice. They are positioned at links between sites $\mathbf{n}=(n_x,n_y)$, as sketched in  \Fig{fig:1}a. On a link at site $\mathbf{n}$, and pointing in direction $i$, the field operators act on electric field eigenstates as $\hat{E}_{\mathbf{n},i}\ket{j}_{\mathbf{n},i} = j\ket{j}_{\mathbf{n},i}$, $\hat{U}_{\mathbf{n},i}\ket{j}_{\mathbf{n},i} = \ket{j+1}_{\mathbf{n},i}$, with $j\in \mathbb{Z}$ \cite{kasper2020jaynes}. Local U(1)-symmetry implies conservation of the Gauß operators  $\hat{G}_\mathbf{n} = \sum_{i=x,y} (\hat{E}_{\mathbf{n},i}-\hat{E}_{\mathbf{n}-\hat{i},i}) - Q_\mathbf{n}/e$, where the $Q_\mathbf{n}$ represent classical static charges, $e$ is the gauge coupling and $\hat{i}$ is the unit vector in~direction~$i$. In temporal-axial gauge, the Hamiltonian reads
\begin{align}
\label{eq:QED-Hamiltonian}
	\hat{H}_{\text{QED}} &= \frac{e^2}{2} \sum_{\mathbf{n}} (\hat{E}_{\mathbf{n},x}^2 + \hat{E}_{\mathbf{n},y}^2) \\
	&- \frac{1}{2a_s^2 e^2}\sum_{\mathbf{n}} \left( \hat{U}_{\mathbf{n},x}\hat{U}_{\mathbf{n}+\hat{x},y}\hat{U}^\dagger_{\mathbf{n}+\hat{y},x}\hat{U}^\dagger_{\mathbf{n},y} + \text{h.c.}\right) \nonumber \; ,
\end{align}
with spatial lattice spacing $a_s$. The first line of \eqref{eq:QED-Hamiltonian} describes the electric energy. The second line represents the magnetic energy, which introduces a gauge invariant coupling of neighboring links in a closed plaquette of the lattice. Electric field operators have been made dimensionless by rescaling with appropriate powers of $a_s$, and we define a dimensionless coupling parameter as~$g^2 = e^2 a_s$.

\textit{Cold-atom implementation.}
We propose to realize a quantum simulator for Hamiltonian (\ref{eq:QED-Hamiltonian}) using an atomic mixture of two spinor Bose gases, trapped in a two-dimensional optical lattice, see \Fig{fig:1}a. For the mapping we associate the gauge links of the original formulation with bosonic annihilation/creation operators, and electric fields with respective number operators. More precisely, 
%The 'physical' lattice is shifted with respect to the links of the lattice gauge theory.
within each well labeled $\mathbf{n}$ according to its lower-left corner, we focus on two spin states for both species $A$ and $B$: $\ket{0}_{\mathbf{n},A/B}$ and $\ket{1}_{\mathbf{n},A/B}$. We then couple neighboring wells through laser-assisted tunneling with appropriately chosen frequencies, which yield the superposition states $\ket{+}_{\mathbf{n},x} = (\ket{0}_{\mathbf{n}-\hat{y},B} + \ket{1}_{\mathbf{n},B})/\sqrt{2} $ and  $\ket{+}_{\mathbf{n},y} = (\ket{0}_{\mathbf{n}-\hat{x},A} + \ket{1}_{\mathbf{n},A})/\sqrt{2} $, see Fig.~\ref{fig:1}a. These superposition states $\ket{+}_{\mathbf{n},i}$ are the ones which are identified with the gauge link degrees of freedom of link $(\mathbf{n},i)$ of the original lattice gauge theory formulation, and the associated creation (annihilation) operator we denote by $\hat{b}^\dagger_{\mathbf{n},i}$~($\hat{b}_{\mathbf{n},i}$).  Hence, the wells of the optical lattice are centered in the associated plaquettes of the lattice gauge theory. The appropriate gauge field dynamics arises then from the on-site homo- and hetero-nuclear contact interactions between the atoms within each well. This construction constitutes the crucial ingredient for our proposed implementation: it allows gauge fields on links to interact with two neighboring plaquettes, thus realizing the gauge-invariant plaquette interaction for magnetic fields via atomic collisions of sizable strength.

The construction yields the cold-atom Hamiltonian
\begin{align}
\label{eq:CA-Hamiltonian}
	\hat{H}_\text{CA} &= \frac{\chi}{2}\sum_{\mathbf{n}}\sum_{i=x,y} \hat{b}^\dagger_{\mathbf{n},i}\hat{b}^\dagger_{\mathbf{n},i}\hat{b}_{\mathbf{n},i}\hat{b}_{\mathbf{n},i} \nonumber \\
	&- \lambda \sum_\mathbf{n} \left( \hat{b}^\dagger_{\mathbf{n},x}\hat{b}^\dagger_{\mathbf{n}+\hat{x},y}\hat{b}_{\mathbf{n}+\hat{y},x} \hat{b}_{\mathbf{n},y}+ \text{h.c.} \right) \nonumber \\
	&+ \frac{\delta}{2}\sum_{\mathbf{n}} (\hat{b}^\dagger_{\mathbf{n},y}\hat{b}_{\mathbf{n},y}+\hat{b}^\dagger_{\mathbf{n}-\hat{y},y}\hat{b}_{\mathbf{n}-\hat{y},y})\nonumber\\
	&\qquad\quad \times(\hat{b}^\dagger_{\mathbf{n},x}\hat{b}_{\mathbf{n},x}+\hat{b}^\dagger_{\mathbf{n}-\hat{x},x}\hat{b}_{\mathbf{n}-\hat{x},x})\; .
\end{align}
Here, the first line encodes intra-species density interactions corresponding to the electric field energy, see~\Fig{fig:1}b. The inter-species collisions in the second line describe the desired magnetic plaquette interaction. There are also additional gauge-invariant interaction terms, described by the third and fourth lines, which represent inter-species density interactions. Direct tunneling is considered to be sufficiently suppressed through an external gradient such that associated processes can be neglected and do not appear in (\ref{eq:CA-Hamiltonian}). Since we are dealing with a lattice field theory, we furthermore exploited that processes of order $\mathcal{O}(a_s)$ vanish in the continuum limit $a_s\rightarrow 0$. We also considered an equal strength of homo-nuclear density interactions in both species. All coupling constants are assumed to be positive $\chi,\lambda,\delta >0$, which depends, in general, on the choice of species. The sign of $\lambda$ can always be chosen by appropriate initialization of the condensate phases. 

%These expressions result in the dimensionless coupling constants  $g^2\bar{n} = \sqrt{\frac{U_{i}( c_1 + 2c_0)}{U_{12}c_{\text{scc}}}} $ and $\chi= \frac{4 U_{12}\tilde{c}_0}{U_{i}( c_1 + 2c_0)}$. Typically one finds $c_1, c_{\text{scc}} \ll c_0,\tilde{c}_0 $, where density scatterings are about one hundred times stronger than the spin interactions. This results in the approximate expressions $\chi \sim  \frac{2 U_{12}}{U_{i}}$ and $ g^2 \sim \frac{20 \sqrt{\chi^{-1}}}{\bar{n}}$.

To explain the relation of our cold-atom Hamiltonian with the Hamiltonian (\ref{eq:QED-Hamiltonian}), we employ a number-phase representation of the bosons, where $\hat{b}^\dagger_{\mathbf{n},i} = \sqrt{\hat{n}_{\mathbf{n},i}} \exp(\mathrm{i}\hat{\phi}_{\mathbf{n},i})$ with commutation relations $[\hat{n}_{\mathbf{m},j},\exp(\mathrm{i}\hat{\phi}_{\mathbf{n},i})]=\exp(\mathrm{i}\hat{\phi}_{\mathbf{n},i}) \delta_{\mathbf{n}\mathbf{m}}\delta_{ij}$. We expand the Hamiltonian \eqref{eq:CA-Hamiltonian} around a large background atom~number~$\bar{n} \gg 1$~and~find 
\begin{align}
\label{eq:QED-limit}
	\hat{H}_\text{CA} = &\,\hat{H}_\text{QED} + \mathcal{O}\left(\frac{\hat{n}-\bar{n}}{\bar{n}}\right) \nonumber\\
	&+ \frac{\delta}{2}\sum_{\mathbf{n}} (\hat{\mathcal{E}}_{\mathbf{n},y}+\hat{\mathcal{E}}_{\mathbf{n}-\hat{y},y})(\hat{\mathcal{E}}_{\mathbf{n},x}+\hat{\mathcal{E}}_{\mathbf{n}-\hat{x},x})\; ,
\end{align}
where we dropped irrelevant constants.
To identify the leading cold atom contribution with compact QED, we represent the atomic electric field operators as deviations of the boson number occupation from the background, $\hat{\mathcal{E}}_{\mathbf{n},i} = \hat{n}_{\mathbf{n},i} - \bar{n}$, and the corresponding gauge links by the approximately unitary operators, $\hat{\mathcal{U}}_{\mathbf{n},i} =\hat{b}^\dagger_{\mathbf{n},i}/\sqrt{\bar{n}} = \exp(\mathrm{i}\hat{\phi}_{\mathbf{n},i})+\mathcal{O}\left(\hat{\mathcal{E}}_{\mathbf{n},i}/\bar{n}\right)$ \cite{yang2016analog,zohar2011confinement,kasamatsu2013atomic}. This identification fulfills the commutation relations of compact QED with the exception of $[\hat{\mathcal{U}}^\dagger_{n,i},\hat{\mathcal{U}}_{m,j}] = 1/\bar{n} \rightarrow 0$
, which becomes accurate in the limit of large background atom numbers~$\bar{n} \gg 1$.
%The situation is similar to that encountered in quantum link models \cite{chandrasekharan1997quantum}, where the gauge field is represented by a spin degree of freedom. [[but here it is better because we use condensates!?]]

While all terms in the effective cold-atom Hamiltonian \eqref{eq:CA-Hamiltonian} are gauge-invariant, the interaction $\sim \delta$ does not lead to a Lorentz scalar in the continuum limit. Thus, for a given lattice size, one has to limit the strength $\delta $ with respect to the electric field energy. Experimentally this may be efficiently achieved by reducing the inter-species overlap with respect to the homo-nuclear one. The overall time scale of the experiment is of the order $\SI{100}{\hertz}$ as recently demonstrated in the experimental setup of Ref.~\cite{mil2020scalable}. In the limit of negligible $\delta$ we then identify the atomic QED coupling constants as $e^2 = \chi$, $a_s^2 = 1/(4\chi \lambda \bar{n}^2) $, which yields the dimensionless coupling $g^2 = \sqrt{\chi/4\lambda \bar{n}^2}$. By appropriate choice of parameters one can in principle tune $g^2$ across a wide range of values reaching both the strong and the weak coupling regime. However, one needs to ensure convergence of physical observables with respect to further increases of $\bar{n}$ while keeping the dimensionless gauge coupling fixed by, e.g., decreasing $\lambda$. Experimentally this can be achieved by reducing the ratio of inter-species to homo-nuclear overlaps in a similar way as for~$\delta$.

\textit{Confinement in compact QED.}
Confinement is easiest understood for strong coupling, where the electric field energy dominates. For two opposite elementary charges it confines their static electric field to a string of unit flux along the shortest path.
For weak couplings, magnetic field interactions tend to delocalize this flux string. While it is destroyed in three-dimensional compact QED, it remains intact in two dimensions: At zero temperature, the two-dimensional theory confines charges in both limits $g^2 \rightarrow \infty$ and $g^2 \rightarrow 0$ \cite{drell1979quantum,polyakov1994compact}. Then the energy density in the flux string between two opposite external charges remains constant with their separation $D$. Thus their combined potential energy increases linearly with $D$ such that the charges are confined.

\begin{figure}
	\flushleft
	\includegraphics[width=0.48\textwidth]{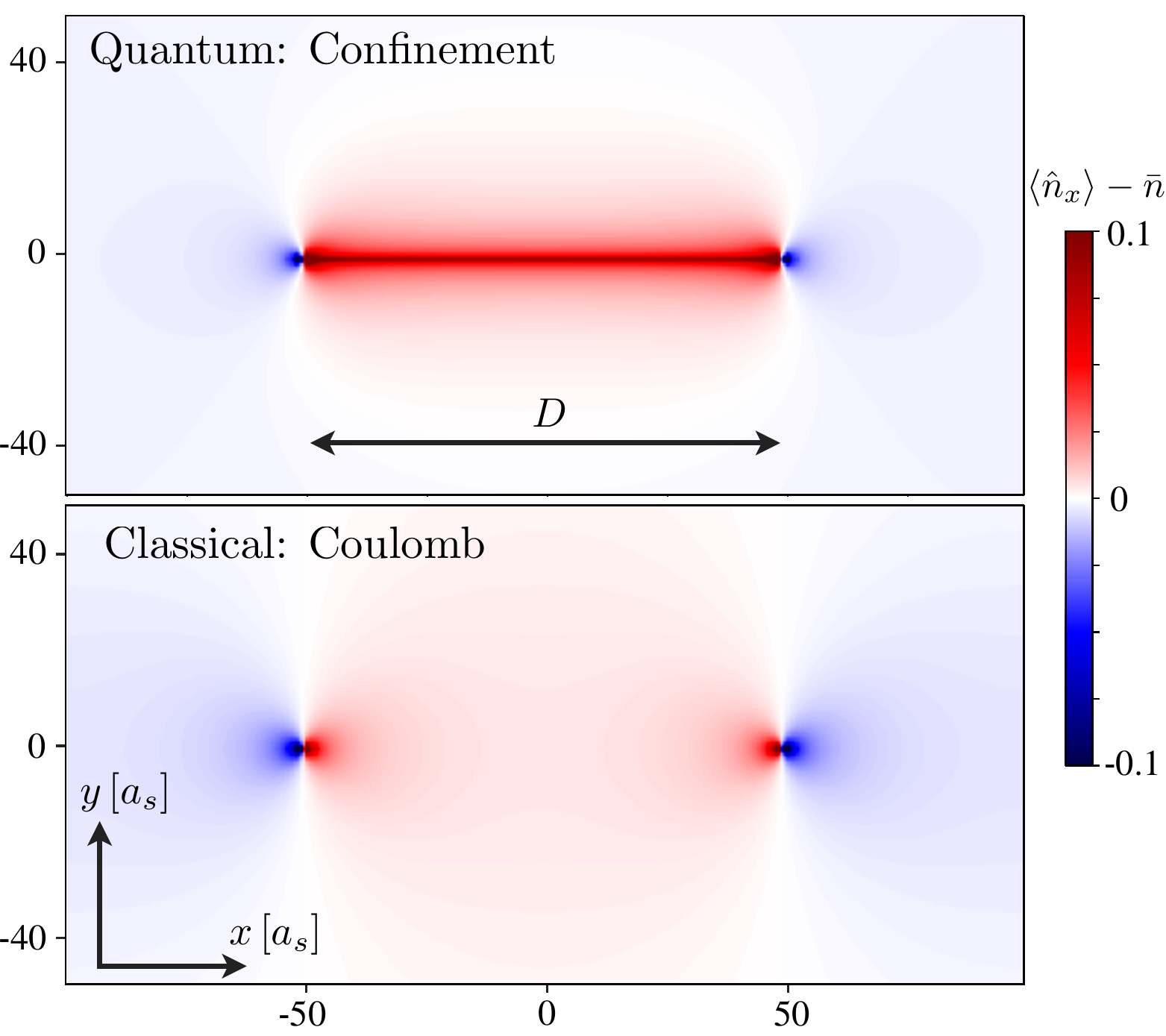}
	\caption{Top: The relative atom number on the $x$-links, $\langle \hat{n}_{\mathbf{n},x} \rangle - \bar{n}$, exhibits a confining string (top) between two opposite external charges in the quantum theory. Here, the distance is $D=100a_s$ with $g^2=1$ and $\mu^2/e^2 \approx 0.071/a_s$. Bottom: The corresponding classical Coulomb case is shown by switching off quantum fluctuations ($g^2 \rightarrow 0$).} 
	\label{fig:2}
\end{figure}
As an illustration of our approach, we characterize confinement in the weak coupling regime employing the variational technique of Ref.~\cite{drell1979quantum}. Choosing a gauge-invariant superposition of quadratic trial wave functionals, one minimizes the Hamiltonian energy density under the Gauß constraint of two external charges with separation~$D$. For large separation~$D$, this procedure yields the energy~$ \langle \hat{H}_{\text{QED}}\rangle \sim \sigma D $, where~$\sigma =  (\pi^2 - 4)\mu^2/2\pi^3 $ is the string tension \footnote{\label{Note1}We have a factor 2 difference compared to Ref.~\cite{drell1979quantum}.} and~$\mu^2a_s^2 = 4\pi^2 \exp(-(2g)^{-2} \int_{-\pi}^{\pi}  \text{d}^2 k/\sqrt{4-2\cos(k_x)-2\cos(k_y)})$ is a characteristic scale, generated by self-interactions of the gauge fields, and setting the length scale of confinement~$D \gtrsim e^2/\mu^2$~\cite{drell1979quantum}.

%For large $D$, this procedure yields the string energy~$ \langle \hat{H}_{\text{QED}}\rangle \sim \sigma D $, with string tension $\sigma$. In compact QED it is given by $\sigma =  (\pi^2 - 4)\mu^2/4\pi^3 $, where~$\mu^2a_s^2 = 4\pi^2 \exp(-(2g)^{-2} \int_{-\pi}^{\pi}  \text{d}^2 k/\sqrt{4-2\cos(k_x)-2\cos(k_y)})$ is a characteristic scale, generated by self-interactions of the gauge fields, which sets the length scale of confinement $D \gtrsim e^2/\mu^2$.

% It can be interpreted as the energy density confined to the flux tube connecting the charges (\Fig{fig:2}).  Regarding the electric field strength of the flux field in the center of \Fig{fig:2}, we find $\langle \hat{E}_{x}(x,0) \rangle\rightsquigarrow -\frac{ \mu^2}{\pi g^2} \log(\frac{ \mu^2}{\pi g^2})$ and $\langle \hat{E}_{y}(x,0) \rangle \rightarrow 0$ for $D\rightarrow \infty$, see Appendix.

\textit{Confinement in a cold-atom quantum simulator.}
We now turn to the cold-atom implementation of compact QED, \eqref{eq:CA-Hamiltonian} and \eqref{eq:QED-limit}, which includes the two main imperfections of a finite boson number and the inter-species density interactions. In the following, we account for both of them using a perturbative analysis around the variational ground state wave functional in the limit $\bar{n} \rightarrow \infty$ and $\delta/\chi \rightarrow 0$ for small couplings $g^2$ to compute atom numbers $\langle \hat{n}_{\mathbf{n},i}\rangle$ and the ground state energy $\langle \hat{H}_\text{CA}\rangle$.

Since the perturbations enter the ground state only in second order, the leading result for the atom number gives the same electric field as in compact QED, $\langle\hat{\mathcal{E}}_{\mathbf{n},i}\rangle = \langle\hat{n}_{\mathbf{n},i} \rangle- \bar{n} =\langle \hat{E}_{\mathbf{n},i}\rangle$. For the atom number on $x$-links in the center of the string we find $\langle \hat{n}_{x} \rangle\approx \bar{n} - \mu^2a_s/(\pi e^2) \log( \mu^2a_s/(\pi e^2))$, for $y$-links we get $\langle \hat{n}_{y} \rangle\approx \bar{n} $ in the limit of infinite volume and charge separation. For finite lattices, which are relevant experimentally, we show the result in Fig.~\ref{fig:2} where one observes that the atoms form a confining flux tube between two external charges at~$\mathbf{n}_+=\left(\text{-}D/\left(2a_s\right),0\right)$ and~$\mathbf{n}_-=(D/(2a_s),0)$. The result is obtained on a $200 \times 200$ square lattice with periodic boundary conditions.
While our theoretical estimates are strictly valid only in the $g^2\ll 1$ region, confinement is expected to persist for all values of the gauge coupling. For the illustration of Fig.~\ref{fig:2} we have evaluated our theoretical results at $g^2=1$, i.e.~extrapolating beyond the well-established weak coupling regime. The validity of such a procedure could be tested in a future quantum simulation experiment at intermediate to strong couplings~\cite{bender2020real}.

\begin{figure}
 	\flushleft
 	\includegraphics[width=0.48\textwidth]{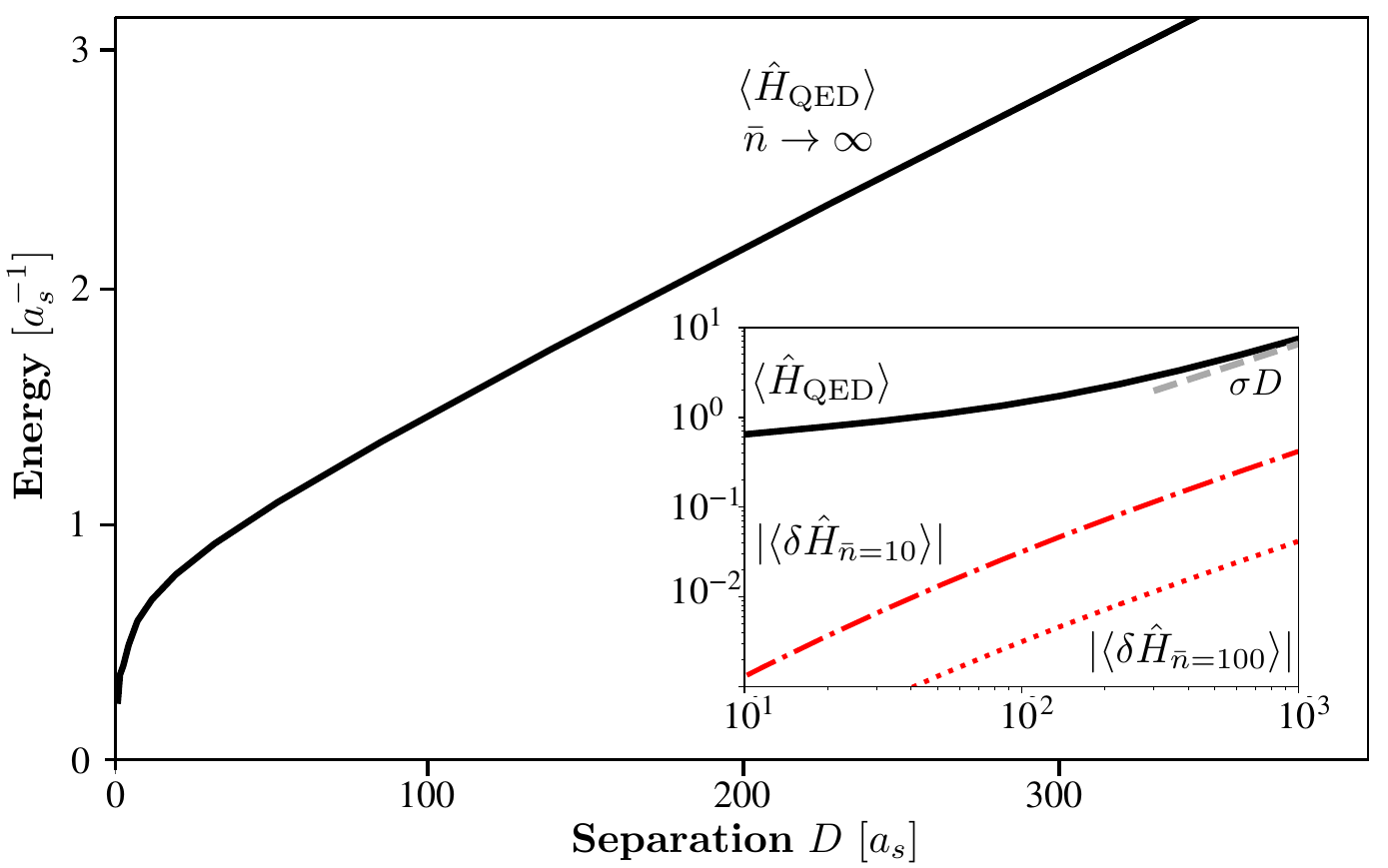}
 	\caption{Ground state energy $\langle \hat{H}_\text{QED}\rangle$ in presence of two external opposite charges at separation $D$. The linear confining potential dominates at large $D$. The inset shows the ground state energy on a logarithmic scale, as well as the leading finite $\bar{n}$ corrections for $\bar{n} =10$ (red dashed-dotted) and  $\bar{n} =100$ (red dotted). The dashed gray line is the analytic asymptotic result from Ref.~\cite{drell1979quantum}, with string tension $\sigma = (\pi^2-4)\mu^2/(2\pi^3)$~\cite{Note1}. }
 	\label{fig:3}
 \end{figure}

The result for the energy expectation $\langle \hat{H} \rangle$ of the ground state is shown in Fig.~\ref{fig:3}. Results are obtained on a $4000\times 4000$ lattice for $g^2=1$, corresponding to $\mu^2/e^2 \approx 0.069/a_s$. While at short distances ($D\lesssim e^2/\mu^2$) the classical Coulomb energy contribution is visible, at large distances  ($D\gg e^2/\mu^2$)  the ground state energy is dominated by the linear confining potential generated from quantum fluctuations. The first correction to the QED Hamiltonian due to the finite boson number of the cold-atom realization (\ref{eq:QED-limit}) is given by $\delta\hat{H}_{\bar{n}} = -(4\bar{n} a_s^2 e^2)^{-1}\sum_\mathbf{n} (\hat{\mathcal{E}}_{\mathbf{n},x}+\hat{\mathcal{E}}_{\mathbf{n},y}+\hat{\mathcal{E}}_{\mathbf{n}+\hat{x},y}+\hat{\mathcal{E}}_{\mathbf{n}+\hat{y},x} + 2)\cos(\hat{\phi}_{\mathbf{n},x}-\hat{\phi}_{\mathbf{n},y}+\hat{\phi}_{\mathbf{n}+\hat{x},y}-\hat{\phi}_{\mathbf{n}+\hat{y},x})$, which close to the continuum limit ($a_s \rightarrow 0$) reduces to~$\delta\hat{H}_{\bar{n}} \rightarrow -(2\bar{n} a_s^2e^2)^{-1} \sum_\mathbf{n} \cos(\hat{\phi}_{\mathbf{n},x}-\hat{\phi}_{\mathbf{n},y}+\hat{\phi}_{\mathbf{n}+\hat{x},y}-\hat{\phi}_{\mathbf{n}+\hat{y},x})$, up to an irrelevant constant. With the identification $\hat{\mathcal{U}}_{\mathbf{n},i} \approx \exp(\mathrm{i}\hat{\phi}_{\mathbf{n},i})$ this term is of the same structure as the plaquette term in~\eqref{eq:QED-Hamiltonian}.
Hence, at leading order, we obtain a correction to the string energy $\langle\delta\hat{H}\rangle_{\bar{n}} =- 2\mu^2 D /(\pi^3\bar{n})$, which smoothly goes to zero for large $\bar{n}$, see the inset of~\Fig{fig:3}.

The second correction concerns the inter-species density interactions with strength $\delta$. We find their contribution vanishes to linear order in $\delta$ within our perturbative analysis. This follows from the symmetry property of the ideal QED ground state (with external charges at $y=0$) under the parity transformation $\mathcal{P}y = -y$: electric fields transform as vectors; they are antisymmetric along $y$-direction. While the leading ($\delta \rightarrow 0$) part of the Hamiltonian is invariant under this transformation, the linear $\delta$-correction is antisymmetric and vanishes. We hence conclude that confinement persists against first order corrections in our proposed implementation scheme.

\textit{Initial state preparation and observables.}
To observe confinement in our 2D quantum simulator we propose to prepare the system in its electric ground state in the limit $g^2 \rightarrow \infty$, corresponding to spatially separated species, i.e. $\lambda \rightarrow 0$. 
We assume that within each well of the optical lattice the atoms initially condense in the spin state $\ket{1}$ for both species. Then, electric field eigenstates are obtained by ramping up the coupling of the respective neighboring clouds to form the symmetric superpositions~$\ket{+}$~\footnote{Alternatively one can choose the antisymmetric state $\ket{-}$ for each species.}.

Introducing a pair of external charges $\pm Q$ requires single-site control over the atom number on a single-particle level. This allows to choose the initial state $\ket{\psi}$ as a product state of electric field eigenstates in accordance with the Gauß sectors $\hat{G}_{\mathbf{n}_{\pm}}  \ket{\psi}=\mp Q/e\ket{\psi}$ and $\hat{G}_{\mathbf{n}}  \ket{\psi}=0$ elsewhere.  \vskip 0.4cm

The two species are slowly brought into spatial contact to adiabatically decrease the coupling $g^2$, where $\ket{\psi}$ stays close to the instantaneous ground state under the Gauß constraint. This adiabatic sweep, see e.g.~Ref.~\cite{zohar2012simulating}, is possible in two-dimensional compact QED which does not exhibit a confinement-deconfinement transition as a function of the coupling \cite{drell1979quantum,polyakov1994compact}. Since its spectrum is gapped for any value of $g^2$, the adiabaticity condition can in principle be upheld deep into the weak coupling region. However, in order to reach weak couplings, long coherence times are required. Throughout this protocol, the flux tube can be observed as shown in \Fig{fig:2} via the local expectation values
$\langle \hat{\mathcal{E}}_{\mathbf{n},x} \rangle = \langle \hat{n}_{\mathbf{n},x}\rangle -\bar{n}$.
The number expectation value $ \langle \hat{n}_{\mathbf{n},x}\rangle$ therein is obtained as the total atom number in the corresponding underlying spin states $\ket{0}_{\mathbf{n}-\hat{y},B}$ and $\ket{1}_{\mathbf{n},B}$.

\textit{Conclusions and outlook.}
In summary, we have proposed a scalable and highly tunable scheme to quantum simulate a two-dimensional gauge theory, including magnetic fields, with the help of spin-changing collisions in an atomic mixture. Our implementation scheme can be applied to general non-equilibrium situations, for instance to study thermalization dynamics in an isolated, interacting gauge theory as a model capturing certain aspects of strongly correlated gauge fields in heavy ion collisions~\cite{berges2020thermalization}. Including fermions would give rise to experimental studies of the string breaking mechanism \cite{hebenstreit2013real}, as well as quantum anomalies out of equilibrium \cite{mueller2016anomaly,ott2020non} or Schwinger pair production beyond one spatial dimension \cite{kasper2014fermion}. Such far-from-equilibrium phenomena typically include high occupations of the gauge fields and are expected to require large $\bar{n}$ \cite{kasper2017implementing}.

We illustrated the physics of confinement in a quantum simulator and outlined an experimental procedure to observe an electric flux tube throughout a broad range of coupling values. In principle, the experimental procedure could directly be applied to the setup of Ref.~\cite{mil2020scalable}, which in view of our results can also be interpreted as a building block of a two-dimensional $U(1)$ gauge theory. Already for a single plaquette it is possible to observe a precursor of confinement via non-perturbative corrections to the classical electric field values \cite{drell1979quantum,zohar2012simulating}. Since this requires high control over local atom numbers, it might be beneficial to  experimentally consider the case of fewer atoms per link first, which may already provide reliable results for large couplings~$g^2$ \cite{zohar2012simulating,bender2020real,haase2020resource,paulson2020towards}. An extension of this work to three dimensions would allow one to study the transition from a deconfined phase at small coupling to confinement at large values \cite{drell1979quantum,polyakov1994compact}. 

% By putting BECs of ultra-cold spinor Bose gases on every link, we approach compact QED in the limit of large atom numbers. We illustrated our setup by showing how interaction-induced charge confinement in the weak coupling regime is represented by the cold-atom system. We showed that this phenomenon is robust against experimental imperfections arising from a finite boson number and inter-species density interactions. 

%To uncover the physics of confinement in a quantum simulator, we proposed an experimental procedure to observe an electric flux tube throughout a broad range of coupling values, starting from $g^2\rightarrow \infty$. This flux tube becomes apparent in local atom number expectation values relative to the background atom number on every link, and can thus be measured straightforwardly.

{\it Acknowledgments.}
We are indebted to Philipp Hauke for valuable discussions and collaborations on related work. We also thank Apoorva Hegde, and Andy Xia for discussions on the experimental parameters. This work is funded by the DFG (German Research Foundation) – Project-ID 27381115 – SFB 1225 ISOQUANT. This work was supported by the Simons Collaboration on UltraQuantum Matter, which is a grant from the Simons Foundation (651440, P.Z.). F.J. acknowledges the Emmy-Noether grant (Project-ID 377616843). \newline
\bibliographystyle{apsrev4-1} 
\bibliography{bibliography}

\end{document}